\documentclass[iop]{emulateapj}
\usepackage{amsmath}
\usepackage{graphicx}
\usepackage{mathtools}
\usepackage[caption=false]{subfig}

\newcommand{\beq}{\begin{equation}}
\newcommand{\eeq}{\end{equation}}
\newcommand{\bea}{\begin{eqnarray}}
\newcommand{\eea}{\end{eqnarray}}
\newcommand{\mrm}{\mathrm}

\newcommand{\dd}{\mrm{d}}
\newcommand{\mdot}{\dot{M}}

\newcommand{\sonic}{{\rm sp}}

\shortauthors{Ro \& Matzner}

\begin{document}

\title{On the Launching and Structure of Radiatively Driven Winds in Wolf-Rayet Stars}

\author{Stephen Ro\email{ro@astro.utoronto.ca} \& Christopher D. Matzner}
\affil{Department of Astronomy \& Astrophysics, University of Toronto, 50 St. George St., Toronto, ON M5S 3H4, Canada}

\begin{abstract}
Hydrostatic models of Wolf-Rayet stars typically contain low-density outer envelopes that inflate the stellar radii by a factor of several and are capped by a denser shell of gas. Inflated envelopes and density inversions are hallmarks of envelopes that become super-Eddington as they cross the iron-group opacity peak, but these features disappear when mass loss is sufficiently rapid.   We re-examine the structures of steady, spherically symmetric wind solutions that cross a sonic point at high optical depth, identifying the physical mechanism by which outflow affects the stellar structure, and provide an improved analytical estimate for the critical mass loss rate above which extended structures are erased.  Weak-flow solutions below this limit resemble hydrostatic stars even in supersonic zones; however, we infer that these fail to successfully launch optically thick winds.  Wolf-Rayet envelopes will therefore likely correspond to the strong, compact solutions.  We also find that wind solutions with negligible gas pressure are stably stratified at and below the sonic point.  This implies that convection is not the source of variability in Wolf-Rayet stars, as has been suggested; but, acoustic instabilities provide an alternative explanation.  Our solutions are limited to high optical depths by our neglect of Doppler enhancements to the opacity, and do not account for acoustic instabilities at high Eddington factors; yet they provide useful insights into Wolf-Rayet stellar structures. 

\end{abstract}


\section{Introduction}
The importance of mass loss in massive stellar evolution is most evident in the Wolf-Rayet (WR) stars, whose defining feature is an optically thick stellar wind. WR winds are an order of magnitude more dense than winds of O-type stars, which is sufficient to  extend the line- and continuum-forming regions into the wind \citep{2007ARA&A..45..177C}. A WR star's wind enshrouds its hydrostatic interior, and hides fundamental stellar parameters such as mass, radius, and rotation from direct observation. 

This is problematic for the study of phenomena that hinge on these parameters, on the detailed stellar structure, or on the star's evolution. Examples include binary evolution, tidal interactions, and WR populations within starburst and Wolf-Rayet galaxies \citep{1999A&AS..136...35S}.   Should a WR star undergo core collapse and explode, the radius and structure of its outer envelope control the production of a shock breakout flash and the pattern of its fast ejecta \citep{1999ApJ...510..379M, 2013ApJ...773...79R} as well as the properties of its early light curve \citep{chevalier92,2010ApJ...725..904N,2011ApJ...728...63R}.

The uncertain regions of WR structure are not small. \cite{2006A&A...457.1015H} and \cite{2006ApJ...636.1033C}  estimate hydrostatic radii ($R_*$) by extrapolating the wind structure to a Rosseland optical depth of $\sim 20$, assuming a $\beta$-law velocity structure \citep{1975ApJ...195..157C}: \[v = v_{\infty}(1- {R_*}/{r} )^{\beta}.\]

Taking $v_\infty$ from observation and fixing $\beta = 1$, these authors infer hydrostatic radii ($\sim 3 - 10 $R$_{\odot}$) up to an order of magnitude larger than those of reference models ($\sim1 $R$_{\odot}$).   
Although the $\beta$-law profile is uncertain, this raises the first question: what inflates WR structures? 

The strongest clue in this puzzle has been the discovery (by the OPAL opacity project: \citealt{1992ApJS...79..507R}) of an opacity peak  at temperatures around $10^{5.2}$\,K due to bound-bound and bound-free transitions of iron nuclei.
The peak, which joins another due to He\,II at $T\sim 10^{4.6-4.8}$\,K,
gained considerable support by resolving the `bump and beat' mass discrepancies in Cephied variable models \citep{1992ApJ...385..685M}. 

In the WR context the Fe and He opacity peaks are especially important, as these stars are not far below the electron-scattering Eddington limit.  The Eddington ratio $\Gamma(r) = \kappa(r) L(r)/[4\pi G M(r) c]$ increases by a factor of several as temperatures cross through the Fe opacity peak, so that $\Gamma$ approaches or even exceeds unity.  \cite{NL02} suggest this to be the root cause of these stars' thick winds, which ​has been supported by wind models from \cite{2005A&A...432..633G}.

Hydrostatic models of WR stars do indeed show inflated envelopes.
\cite{1999PASJ...51..417I}, \cite{petrovic}, and \cite{grafener} construct such models using updated OPAL opacity tables, and discover a significant redistribution of stellar material due to the Fe opacity bump.  
In these regions $\Gamma$ approaches unity, and the pressure becomes strongly dominated by radiation because ${d P_{\rm gas}(r)}/{d P_{\rm rad}(r)} = 1/\Gamma(r) - 1$ (in hydrostatic, radiative zones). The density scale height can also become very large, as it scales inversely with the local effective gravity $g_{\rm eff}=(1-\Gamma)GM(r)/r^2$.   Gas density therefore declines only slowly with radius,  a feature which is not erased by the onset of convection.  (Envelope inflation has also been observed in non-WR massive stellar evolution models: \citealt{2015A&A...573A..71K, 2015A&A...580A..20S}.)

A curious structure arises within one-dimensional hydrostatic models where $\Gamma>1$. To balance the net outward force of radiation and gravity,  gas pressure must rise towards the surface.  For this reason, inflated envelope models experience a density inversion and are capped with a denser shell of gas.   While the validity of such structures has been defended \citep{joss}, strong instability is observed in one-dimensional evolutionary models \citep{2013ApJS..208....4P}.  Non-adiabatic stability analyses  \citep{1993MNRAS.262L...7G,2002MNRAS.337..743G,1998MNRAS.294..622S} find extended envelopes  to be excited by violent `strange mode' pulsations. It is very likely that non-hydrostatic, non-steady, or three-dimensional effects arise; \cite{1992iesh.conf..138M} considers strong turbulent motion, mechanical wave luminosity, eruptive geysers, and outflows as plausible scenarios. 

It is important to note that extended outer envelopes in hydrostatic models are often inconsistent with the mass outflow rates of WR stars.   The assumptions of a hydrostatic model are valid only where outflow motions are subsonic and carry a negligible fraction of the luminosity.  The extended envelope of \citet{petrovic}'s models reach densities of $\rho(r)\simeq 10^{-10}$\,g\,cm$^{-3}$ and isothermal sound speeds $c_i(r)\simeq 35$\,km\,s$^{-1}$ at radii $\sim 4 $R$_\odot$.  The outflow Mach number  $v/c_i= \dot M/(4\pi r^2 \rho c_i)$ therefore exceeds unity for $\dot M \gtrsim 10^{-5.4} $M$_\odot$/yr. This is a low value for WR mass loss ($\dot M\sim 10^{-5.5}$ to $10^{-4.0} $M$_\odot$/yr; \citealt{2000A&A...360..227N}).   We therefore have a second question: how much mass loss alter the structure of WR envelopes?

The question is not new. \cite{1992ApJ...394..305K} suggested that an enhanced opacity bump can generate an optically-thick wind and extend the effective photospheric radius. \citeauthor{1992ApJ...394..305K} constructed an artificial opacity peak to test this hypothesis, and discovered a core-halo configuration in which the original compact stellar core is surrounded by an optically-thick outflow. 

\cite{petrovic} propose an answer based on models in which mass loss is implemented within a stellar hydrodynamic code. Above a critical mass loss rate, which they identify with an outflow speed equal to the escape speed (as estimated from the hydrostatic density profile), extended model envelopes disappear.  However, we are left with several questions. The escape speed is of order $30c_i$; so, why is the structure not altered by mass loss that is thirty times weaker? Second, what special conditions arise at the wind sonic point? Lastly, we are confused by the statement by \citet{petrovic} that they remove ``a proportionate amount of mass from each shell'' in the outer 40\% of the stellar mass, as it is not clear how this corresponds to a steady outflow. A self-consistent model must include the dynamics of the transition between the envelope and the wind. 

Our goal, therefore, is to evaluate the impact of the opacity bump on dynamically self-consistent Wolf-Rayet stellar structures and winds, and to re-examine the consequences of mass loss for the survival of extended envelopes.

Using OPAL opacities, \cite{NL02} analyzed the opacity bump for its capacity to launch a transonic wind. They show that, within a radiation-dominated wind with diminishing radiative luminosity $dL_r/dr<0$,  the sonic point (where $v=c_i$) must reside where the opacity increases outward.  Their examination of the sonic point conditions showed the iron opacity bump and the smaller He\,II bump both have the capacity to launch optically-thick winds and to explain the observed mass loss rates of WR stars. But if opacity bumps are responsible both for envelope inflation and for wind launching, can envelope inflation ever coexist with a wind?

We aim to address this question by solving for the dynamical transition between envelope and wind implied by the iron opacity bump.  We will capitalize on the high optical depths of WR winds by using tabulated Rosseland opacities to integrate through the wind sonic point; this is both a useful simplification, and a limitation of our results.  In \S~\ref{SS:acoustic-instabilities}, we discuss how acoustic instablities can generate density fluctuations, which can modify the effective opacity and alter the stellar structure. For simplicity, we do not include these effects. In \S~\ref{sec:wind}, we describe the assumptions and numerical methods used to construct our WR wind models, and re-examine the sonic point conditions.  In \S~\ref{sec:results} we investigate a range of wind models for the same WR progenitors used by \cite{grafener}, in order to understand the influence of dynamics upon an inflated envelope and to explicitly determine the maximal mass loss rate to retain such a structure. In \S~\ref{SS:Various-masses} we present wind models for a range of progenitor masses. 

\section{Stellar Wind Models}
\label{sec:wind}
We will explore steady, one-dimensional, spherically symmetric models of Wolf-Rayet winds.  We begin by enumerating the equations to be solved (\S~\ref{subsec:Wind_equations}) and then re-examine the sonic point conditions. Therefore, our models become inaccurate where the spherical flow is unstable and where the Rosseland approximation is not appropriate.  Many WR winds are sufficiently optically thick that this approximation is valid through the sonic point.  However, because we do not account for the increase in opacity due to Doppler shifting of the lines, we do not integrate through the wind photosphere.  We cannot, therefore, solve self-consistently for the mass outflow rate.  Instead we adopt a range of values for $\dot M$ and study the structure of the outer envelope and deep wind for each value. 
A list of conditions is discussed in Section \ref{sec:diff}. 

\subsection{Structure equations} \label{subsec:Wind_equations} 
The total pressure $P=P_g+P_r$ is composed of an ideal gas pressure $P_g = \rho k_BT/\mu =\rho c_i^2$ and radiation pressure $P_r=a_rT^4/3 = \rho c_r^2$, where $T$ is temperature, $\mu$ the mean molecular weight (units of mass), and $a_r$ the Stefan-Boltzmann radiation constant. 
Note that $c_i$  is the isothermal sound speed.

We solve a simple set of equations for steady spherical flow, equivalent to those adopted by \citet{NL02}:  mass conservation 
\beq
\frac{\dd}{\dd r}\left( r^2 \rho v \right) = 0,
\label{erho}
\eeq
corresponding to a constant mass loss rate  $\dot M = 4\pi r^2 \rho v$ 
(where $\rho$ is the density, $v$ is the velocity at radius $r$ from the stellar centre); momentum conservation, in the form of the Euler equation 
\beq
v \frac{\dd v}{\dd r} + \frac{1}{\rho}\frac{\dd P}{\dd r} = - \frac{GM}{r^2}; 
\label{emom}
\eeq
and energy conservation,
\beq
L_r +{\cal B} \mdot    = \dot{E} = \mrm{constant},
\label{enrg}
\eeq
where $L_r$ is the radiative luminosity in the fluid frame at radius $r$,
\beq \label{e_BernoulliFactor} 
{\cal B} = w + \frac{1}{2}v^2-v_k^2
\eeq
is the Bernoulli factor, or ratio of energy flux to mass flux, 
$v_k^2=GM/r$ is the negative gravitational potential (square of the Kepler speed), $w=5c_i^2/2+ 4 c_r^2$ is the specific enthalpy, and $\dot{E}$ is the energy loss rate (not including rest energy). 

 We make several approximations in addition to the assumption of steady spherical flow.  
First, we consider only an outer region of negligible mass, so we approximate the enclosed mass with the total stellar mass, $M(r) \simeq M_*$.   Second, as we concentrate on optically-thick regions without appreciable convective luminosity, we employ the radiation diffusion approximation 
\beq
\frac{\dd P_r}{\dd r} = -\kappa \rho  \frac{L_r}{4\pi r^2 c}
\label{ediff}
\eeq
where $\kappa$ is the effective opacity.  Rewriting this in a convenient form, 
\bea
\frac{1}{\rho}\frac{\dd P_r}{\dd \ln r}  
= -\Gamma_r v_k^2,
\label{erad}
\eea
where 
\beq
\Gamma_r = \frac{\kappa L_r}{4 \pi c GM}
\label{eedd}
\eeq
is the local Eddington ratio. In our wind structure calculations we shall employ the Rosseland approximation $\kappa=\kappa_R$ and use tabulated values of $\kappa_R$ from the OPAL project. 

\subsection{Sonic point criteria} \label{subsec:Sonic_Point}
The momentum equation contains a critical point, which supplies several constraints on the behaviour of the wind. These have already been discussed by \citet{NL02}, but we re-examine them to make a couple additional points.  We  substitute the pressure gradient $\dd P/ \dd r=\dd P_g/ \dd r + \dd P_r/ \dd r$ in equation (\ref{emom}) and evaluate this using the temperature gradient implied by equation (\ref{erad}) and the density gradient from equation  (\ref{erho}).  We find 
\beq
v' =\frac{2c_i^2 - v_k^2\left[1-\Gamma_r \left(1+\phi\right)\right]}{v^2-c_i^2},
\label{evprime}
\eeq
where $\phi\equiv P_g/(4P_r)$. (In terms of the more familiar quantity $\beta\equiv P_g/P$, $\phi = \beta/[4(1-\beta)]$.)  Here and elsewhere, a prime indicates a logarithmic derivative with respect to radius, e.g.\ $v' = \dd \ln v/\dd \ln r$.

The critical point is the isothermal sonic point $R_\sonic$, where $v(R_\sonic) = c_i(R_\sonic)$, so that the denominator vanishes in equation (\ref{evprime}).   (We denote sonic-point values with the subscript $\sonic$.) 
For $v'_\sonic$ to be defined, the numerator must also vanish; this shows that the sonic point can {only} exist where the radiative luminosity is sub-Eddington relative to the matter: 
\beq \label{e_Gamma_sp} 
\Gamma_{r, \sonic} = \frac{1 - q_i}{1 + \phi} < 1
\eeq
at $r=R_\sonic$, and this condition applies to accretion as well as outflow.  Here, we define $q_i \equiv 2c_i^2/v_k^2$, and likewise $q_r \equiv 2c_r^2/v_k^2$ for upcoming derivations. 
In WR winds $\Gamma_{r, \sonic}$ is only slightly below unity (cf.\ \citealt{NL02} eq.~40), because $q_i\ll1$ and $\phi\ll 1$.  

In fact, the value of $q_i$ is restricted by the fact that $v_k$ reflects the stellar central temperature, which is moderated by the burning stage, and the fact that $c_i$ is determined by the temperature of the opacity peak.  Evaluating $v_k$ using the  mass-radius relation of \citet{1992A&A...263..129S}, which implies $v_k \simeq 1900 [M/(30\,$M$_\odot)]^{0.21}$\,km\,s$^{-1}$, gives $q_i  \simeq 10^{-3.25} (T/10^{5.2}\,K ) (30\,$M$_\odot/ M ) ^{0.42}$.   However, in real WR stars the sonic point forms at a somewhat larger radius, so that $q_i$ can be a couple times larger than this estimate. 

The velocity gradient at the sonic point must be determined by l'H\^opital's rule, as the ratio of derivatives of the numerator and denominator of equation  (\ref{evprime}).   
Following \citet{NL02}, we note that the denominator increases through the sonic point, and therefore the numerator must as well in order for the wind to accelerate outward ($v'>0$).   Using equation (\ref{e_Gamma_sp}), the radial derivative of the numerator is 
\[2\frac{\dd c_i^2}{\dd r} - q_i \frac{\dd v_k^2}{\dd r} + v_k^2 (1+\phi) \frac{\dd \Gamma_r}{\dd r} + v_k^2 \Gamma_r \frac{\dd \phi}{\dd r}. \] 
The first term can be evaluated with $\dd c_i^2/\dd r = -\Gamma_r \phi v_k^2/r$ (from eq.~\ref{erad}), and combined with the second term, using $\dd v_k^2/\dd r = - v_k^2/r$.   Using equation (\ref{e_Gamma_sp}) a second time, d(numerator)/d$r$ becomes 
\[ v_k^2 \left[ {q_i-(2-3q_i)\phi \over  (1+\phi) r} + (1+\phi) \frac{\dd \Gamma_r}{\dd r} + \Gamma_r \frac{\dd \phi}{\dd r} \right]. \]

The first term is small in magnitude, and negative if $q_i/(2\phi) = 4 c_r^2/v_k^2 <1$.  We note that the Bernoulli parameter $\cal B$ is approximately $4c_r^2-v_k^2$ at the sonic point. Therefore, for the first term to be negative, the wind must be formally bound in the sense of having a negative Bernoulli parameter.  

The second term tends to be negative if $\kappa$ is constant, because $L_r$ tends to decline outward as energy is converted to kinetic form.  On the other hand, this term can be large and positive if $\kappa$ increases sharply outward. 

The last term is negative if $\dd\phi/\dd r <0 $, i.e. when $\rho T^{-3}$ decreases outward.  Note, however, that when a radiation-dominated gas is stable against convection, $\rho T^{-3}$ must decrease outward.  Therefore, this term is negative in a stably stratified wind. 

Our analysis therefore corroborates \citeauthor{NL02}'s conclusion that the wind sonic point is almost certainly located where $d\kappa/dr>0$ so that $\Gamma_r$ is increasing outward.  Combined with the fact that $\Gamma_r$ is only slightly below unity at the sonic point, it is highly likely that the flow will be super-Eddington for some range of radii immediately outside the sonic radius. 

While neither of these statements is absolute, we see that the sonic-point condition in a wind model is essentially identical to the condition for density inversion in a hydrostatic model: namely, that $\Gamma_r$ increase through unity.  We hypothesize that density inversions are {\em always} erased by dynamical winds, and test this later with numerical models.

Our numerical solutions require a quantitative description of the sonic point, which we gain by using the partial derivatives of $\kappa(\rho,T)$, assuming $\kappa$ does not depend on the velocity gradient.
Defining ${k_\rho}\equiv \partial\ln\kappa/ \partial\ln\rho$ and ${k_T}\equiv \partial\ln\kappa / \partial 
\ln T$, we find that $v_{\sonic}'$ satisfies a quadratic equation:
\beq
0 = (v_\sonic')^2 + B v_\sonic' + C,
\label{evprimesp}
\eeq
where
\bea
B &=&  2\Psi + {k_\rho}{\cal W}_i+ 4q_r\xi {\cal W}_i(1+\phi)  \label{evprimesp_b}  
\eea 
and 
\bea
C &=& -6\Psi^2 + 4\Psi - 1 + 2{\cal W}_i\left({k_\rho}+{k_T}\Psi\right) \nonumber \\ 
& \ & \ + \xi {\cal W}_i(3q_i\Psi + 8q_i+8q_r -6) \label{evprimesp_c},
\eea
with the following definitions:
\begin{equation*}
\xi = \frac{\mdot v_k^2}{2L_r},~~
{\cal W}_i = \frac{1-q_i}{q_i},~~
\Psi = \frac{\phi }{1+\phi}{\cal W}_i,
\end{equation*}
all evaluated at the sonic point. 
We see that the solutions depend only on the local properties of the flow and opacity gradients. 

The roots of equation (\ref{evprimesp}) are of the form $v_{\sonic}' = ({\pm \sqrt{B^2-4C}-B })/{2}$,
and equation (\ref{evprimesp_b}) shows that $B>0$.    Real solutions require $4C\leq B^2$; if $C>0$ then both solutions are negative, whereas if $C<0$ then there exists one positive and one negative solution.   We are primarily interested in winds that accelerate outward, i.e., those for which $v_\sonic'\geq0$; this requires that $C\leq0$ and that 
\beq
 v_{\sonic}' = \sqrt{(B/2)^2-C} - B/2. 
\eeq

\subsection{Inner boundary: matching a hydrostatic star} 
Rather than solving for the structures of the wind and star simultaneously, we identify the base of our wind model with conditions at a matching radius within a hydrostatic model.
The exact boundary location is selected to satisfy the following conditions:
\begin{enumerate}
\item The total wind mass is negligible in comparison to the stellar mass;
\item The stellar model is locally chemically homogeneous, $\nabla \mu=0$; and 
\item The flow speed in the wind model is much less than the gas sound speed, $v\ll c_i$.
\end{enumerate}
We have had no difficulty identifying radii at which all these conditions are met. 

We investigate hydrogen-free, chemically homogeneous winds composed of pure helium with solar metallicity $Z = Z_{\odot}=0.02$, and consider a range of stellar masses $M_*=(15, 20, 23,  25, 30) $M$_{\odot}$ are considered to study the phenomena of envelope inflation and mass loss.  We focus in particular on the $23 {M}_{\odot}$ case presented by \cite{grafener}. 

\subsection{Regime of validity}
\label{sec:diff}
In order for our solutions to be valid, several requirements must be met.  First, the flow must be optically thick so that the diffusion approximation is valid; but this is essentially guaranteed in WR winds, so we ignore this constraint.  Second, force enhancement due to the Doppler shifting of spectral lines \citep[e.g.][hereafter CAK]{1975ApJ...195..157C} must not invalidate our use of the Rosseland opacities from the OPAL project.  \citet{NL02} have previously argued that the enhancement is negligible at the wind sonic point, but we revisit the issue throughout our solutions. Third, the subsonic portion of the flow must be stable against convection; or, if convection sets in, it (and any waves it launches) must be too weak to alter the radiative flux.  Fourth, any other instabilities of radiation-dominated fluids \citep[e.g.][]{2003ApJ...596..509B} must also not invalidate the assumption of smooth spherical flow. These instabilities provide additional line broadening and may enhance wind acceleration. Our approach will be to obtain solutions assuming these conditions are met, and then check their validity after the fact.

The Rosseland approximation degrades once absorption lines in the accelerating wind are Doppler-shifted beyond a thermal line-width  across a photon mean-free-path. This occurs \citep{NL02} where the CAK optical depth parameter
\begin{equation}
t_{\mrm{CAK}} = \frac{\sigma_{\mrm{e}}^{\mrm{ref}} v_{\mrm{th}} \rho}{\dd v / \dd r}
\end{equation}
falls below unity, where $\sigma_{\rm e}^{\rm ref}=0.325$\,cm$^2$\,g$^{-1}$ is a reference electron scattering opacity and $v_{\rm th} = 0.8c_i$ is the thermal velocity of protons. We use this criterion to highlight where the force enhancement due to line shifting is likely to be a significant correction. 

Being non-rotating and homogeneous in composition, our flows are unstable to convection where low-entropy matter lies above high-entropy matter according to the sense of the  total acceleration (including gravity), i.e. when 
\beq\begin{array}{lr}
 \nabla_{\mrm{rad}} \ge \nabla_{\rm ad}
 \end{array} 
\label{e_ConvInstab}
\eeq
where $\nabla_{\mrm{rad}} = d\ln T/d\ln P$ is the radiative temperature gradient and `ad' means the adiabatic gradient. The outwardly accelerating flow $\dd v/\dd r >0$ enhances the total acceleration and stabilizes the flow against convection. Further discussion is found in  Section \ref{sec:inflation}.

Finally, we evaluate the growth rates of modes identified by \citet{2003ApJ...596..509B}. These modes are radiation hydrodynamic instabilities, distinct from convection.

\subsection{Numerical Method}
The subsonic region of our flow satisfies a two-point boundary value problem,  between an inner matching location and the sonic point. 
Once the radius of the sonic point is found, the supersonic region is
solved separately as an initial value problem. Our notation and numerical methods are in close accordance to the models of hot Jupiter outflows by \cite{2009ApJ...693...23M}. 

\subsubsection{Subsonic Region: Relaxation Method}
In our work, we use the relaxation solver \texttt{solvede} from Numerical Recipes \citep{1992nrca.book.....P}. This routine interprets the system of differential equations as a multivariate root-finding problem, and requires equations to be in finite-difference (FD) form 
\beq
0 = E_{ij} \equiv \Delta_jy_i - \frac{\dd y_i}{dx}\Delta_jx,\nonumber
\eeq
where $y_i$ are the $i$-th fluid variables and $\Delta_jx\equiv x_j - x_{j-1}$ at the $j$-th grid point. 

The corresponding FD forms of equation (\ref{erho}), (\ref{evprime}), and (\ref{ediff}) are
\bea
E_{1j} &\equiv& \Delta_j\rho - \frac{\dd \rho}{\dd r}\Delta_jr \nonumber \\ 
&=& \Delta_j\rho + \frac{\rho}{r} \left( 2 + v'\right)\Delta_jr,
\eea

\bea
E_{2j} &\equiv& \Delta_jT - \frac{\dd T}{\dd r}\Delta_jr \nonumber \\ 
&=& \Delta_jT +  \left[\frac{3\kappa%_{\mrm{OPAL}}
(\rho,T) \rho L_r}{16 \pi a c r^2 T^3}\right]\Delta_jr,
\eea
and
\bea
E_{3j} &\equiv& \Delta_jv - \frac{\dd v}{\dd r}\Delta_jr \nonumber \\ 
&=& \Delta_jv 
%\nonumber \\ \ \ \ \ \ \ \  \ & \ & 
-  \left\{ \frac{2c_i^2 - v_k^2\left[1-\Gamma_r \left(1+\phi\right)\right]}{v^2-c_i^2}\right\}\frac{v\Delta_jr}{r}.
\eea
The Rosseland opacity $\kappa$ is supplied by the OPAL opacity tables of \cite{1996ApJ...464..943I}.

The current system of equations cannot be solved without the location of the outer boundary or sonic point radius $r = R_{sp}$. The power of the relaxation method is its ability to treat the outer boundary location as a dependent variable, using the definition
\beq
z\equiv R_{sp} - R_*,
\eeq
where $R_*$ is the inner boundary radius, which can be solved for simultaneously. Since there is only one outer boundary, $z$ is a constant. We add the trivial FD equation 
\bea
E_{4j} &\equiv& \Delta_jz = 0.
\eea
We define the new independent spatial variable $q \in [0,1]$ and substitute all instances of radius, 
\beq
r =R_{*}+z q.
\eeq

The four dependent variables $\rho, T, v, z$ are normalized (and non-dimensionalized) to the following fiducial set to maximize numerical precision: $\rho_0 = 10^{-7}$g cm$^{-3}$, $T_0=10^6$ K, $v_0 = 10^7$ cm s$^{-1}$, $z_0=1 $R$_{\odot}$. Within \texttt{solvede}, the convergence parameter \texttt{conv} is set to $10^{-7}$ with the following weighting parameters or \texttt{scalv} used in the error measure: $\rho$: 10;  $T$: 5; $v$: 1; $z$: 1.

The relaxation method is a multidimensional extension of Newton's method, which estimates a set of first-order corrections to the FD equations. This requires partial derivatives of FD equations with respect to dependent variables $\partial E_{ij} / \partial y_i, \ \forall \  i,j$. We compute this with the differentiation package from \texttt{GNU Scientific Library} \citep{Gough:2009:GSL:1538674}. We explicitly use the opacity gradients $ \partial \kappa / \partial \rho $, $ \partial \kappa / \partial T$ supplied by the OPAL opacity tables in all calculations.

\subsubsection{Stellar Parameters and Boundary Conditions}
\label{sec:bc}
The stellar wind requires four local boundary conditions as there are four dependent variables $(\rho, T, v, z)(q)$. The boundary conditions are written in FD form and can conveniently be defined implicitly: we denote them $B_1$ through $B_4$, all of which equal zero when the boundary conditions are satisfied. The sonic point criteria supply two outer boundary conditions at $q = 1$: 
\beq
B_1 \equiv v^2 - \frac{k_B T}{\mu },
\eeq
and
\bea
B_2 &\equiv& 1 - \frac{2c_i^2}{v_k^2} -\Gamma_r \left(1+\phi\right) \nonumber \\
&=&  1 - \left[\frac{2k_B T(R_*+z)}{G M_*\mu }\right] \nonumber \\
& \ & \ \ - \left[\frac{\kappa(\rho,T) L_r}{4 \pi c GM_*}\right] \left(1+\frac{3k_B\rho}{4a\mu T^3}\right) 
\eea

The remaining two inner boundary conditions connect the stellar wind to the hydrostatic interior. The solution across the boundary cannot be definitively smooth nor continuous, since one domain is hydrostatic. However, the approximation becomes very good where the velocity at the base of the wind is small. 

We choose the temperature to be continuous across the boundary and define 
\beq
B_3 \equiv T - T_0.
\eeq
The temperature gradient cannot be continuous across the boundary unless the density and diffusive luminosity are as well. We adjust the diffusive luminosity such that it remains smooth across the boundary. In hydrostatic equilibrium ($\mdot=0$), the diffusive luminosity is effectively unchanged beyond the regions of nuclear fusion and convection $L_{\mrm{core}} = L_{\mrm{rad}} = \dot{E}$. In a wind the diffusive luminosity from the core must be reduced to accelerate and lift the gas out of the potential well. For a WR star, lifting the gas out of the potential is the dominant source of energy lost; the remaining terms are negligible in comparison. Thus, from equation (\ref{enrg}) we have
\bea
L_r &=& \dot{E} - \mdot\left(w + \frac{1}{2}v^2-v_k^2\right) \nonumber\\
&\simeq& \dot{E} + \mdot \frac{GM_*}{R_*}
\label{enrg2}
\eea

The density is constrained implicitly with the mass continuity equation (eq.\ \ref{erho}), 
\beq
B_4 \equiv \mdot - 4\pi R_*^2 \rho v.
\eeq

In summary, the stellar parameters are the luminosity $L_*$, mass $M_*$, mass loss rate $\mdot$, temperature $T_0$, and molecular weight $\mu$ at the base of the wind $R=R_*$. Details of the chemical abundances and metallicity are only necessary in selecting the appropriate OPAL opacity tables.

\subsubsection{Numerical Sonic Point Treatment}
Any numerical method that does not respect a critical point is fortunate to converge at all, let alone be accurate. Simply increasing the resolution is self-defeating because the numerator $N(x_1)$ or denominator $D(x_2)$ need not be zero at the same grid point (ie. $x_1\neq x_2$) or vanish at all! This is disastrous for iterative schemes as the differential equations
may either explode towards both positive and negative infinities, become zero (ie. a breeze solution), oscillate in sign, or any combination of these. 

We evaluate $v'$ in the manner of \cite{2009ApJ...693...23M}:
\beq
\frac{\dd v}{\dd r} = F_{\mrm{exact}}\frac{\dd v}{dr}\Big|_{\mrm{exact}} + (1-F_{\mrm{exact}})\frac{v v_+'}{r},
\eeq
where $\frac{\dd v}{\dd r}\Big|_{\mrm{exact}}$ is given by equation (\ref{evprime}), $v_+'$ is the positive root from equation (\ref{evprimesp}), and
\beq
F_{\mrm{exact}} \equiv -\mrm{erf}\left[ p\left(1-\frac{c_i^2}{v^2}\right) \right],
\eeq
where erf is the error function, and $p=100$ is the transition width for the sonic point. 

\subsubsection{Generating the First Wind Model}
A `good' initial guess for the entire subsonic wind structure is required for the relaxation method to begin. The resulting solution can then be used as an initial guess for the next problem bearing a different set of conditions. 

We found this step to be very challenging, but eventually developed a viable strategy in which we first
solve a trivial problem and successively add additional physical terms. 
We begin with an isothermal wind with critical point $R_\sonic = \frac{G M_*}{2c_i^2}$, and impose the analytical solution \citep{2004AmJPh..72.1397C}. We then allow the temperature to vary, and adjust the constant diffusive luminosity $L_*$ and opacity $\kappa_0$ until they are similar to the conditions within a WR star. Third, we allow $\kappa$ to vary linearly with radius as $\kappa(q) =\kappa_0 + (\kappa_1-\kappa_0)q$, and we vary the opacity limits as we include terms from the Bernoulli factor ${\cal B}$. Finally, we transform from the artificial opacity $\kappa(q)$ to the OPAL tables for $\kappa(\rho, T)$. Once the first wind model is available, generation of subsequent wind models become trivially accessible.

\subsubsection{Supersonic Region: Initial Value Problem}
Since all of the wind variables are defined at the sonic point, we compute the supersonic component as an initial value problem. We use the Bulirsch-Stoer routine contained in the integration module \texttt{odeint} from Numerical Recipes \citep{1992nrca.book.....P}. The convergence parameter is set to $\mrm{EPS}=10^{-13}$. ​Integration is continued until either the flow becomes subsonic or approaches the next partial ionization zone of helium. We find that wind solutions that become subsonic do so for temperatures well above $7 \times 10^4$ K. The remaining solutions are fast and certainly break the Rosseland approximation by this point.

Because we cannot solve for the region of low optical depth in which Doppler-enhanced line forces are significant, we cannot integrate to infinite radius, and we cannot choose a self-consistent value for $\dot M$.   Nevertheless, we can explore wind and envelope structures across a range of mass loss rates.  Often our solutions cross through the opacity peak but fail to accelerate to speeds above the escape velocity, and then decelerate and stall at some radius. If this occurs within the regime of validity of the Rosseland approximation, it indicates a physically inconsistent solution. However, if the Rosseland approximation fails at the radii for which the wind solution stalls, it is possible that line forces would have permitted the wind to escape.

\section{Results}
\label{sec:results}
\subsection{23 $\mrm{M}_{\odot}$ Helium Star}
We use Modules for Experiments in Stellar Astrophysics (\texttt{MESA}, \citealt{2011ApJS..192....3P, 2013ApJS..208....4P}) to construct a 23 M$_{\odot}$ pure helium star with solar metallicity. The remaining 
parameters to define the stellar wind problem are the luminosity $L_*=10^{5.80}\ \mrm{L}_{\odot}$ and temperature $T_0$ at the base of the wind $R_0$. The location of the inner boundary is chosen where log $T_0/$K $=6.0$, well beneath the sonic point and the iron opacity bump. The radius of this boundary is $R_{*}\sim1.44 \ \mrm{R}_{\odot}$. 

Since our system is not hydrostatic, we artificially adjust the stellar luminosity $L_*$ at the base of the wind such that the diffusive luminosity $L_r$ of the wind matches that from MESA. The correction here is about 1 L$_{\odot}$. A discrepancy between the hydrostatic and wind ​density is found of order $(\rho_0 - \rho_{\mrm{MESA}})/\rho_0 \simeq ​10^{-5}$ at the inner boundary. The discrepancy diminishes exponentially towards the interior, as ​expected by \cite{1999isw..book.....L}.

\begin{figure}[h]
    \centering
    \includegraphics[width=\columnwidth]{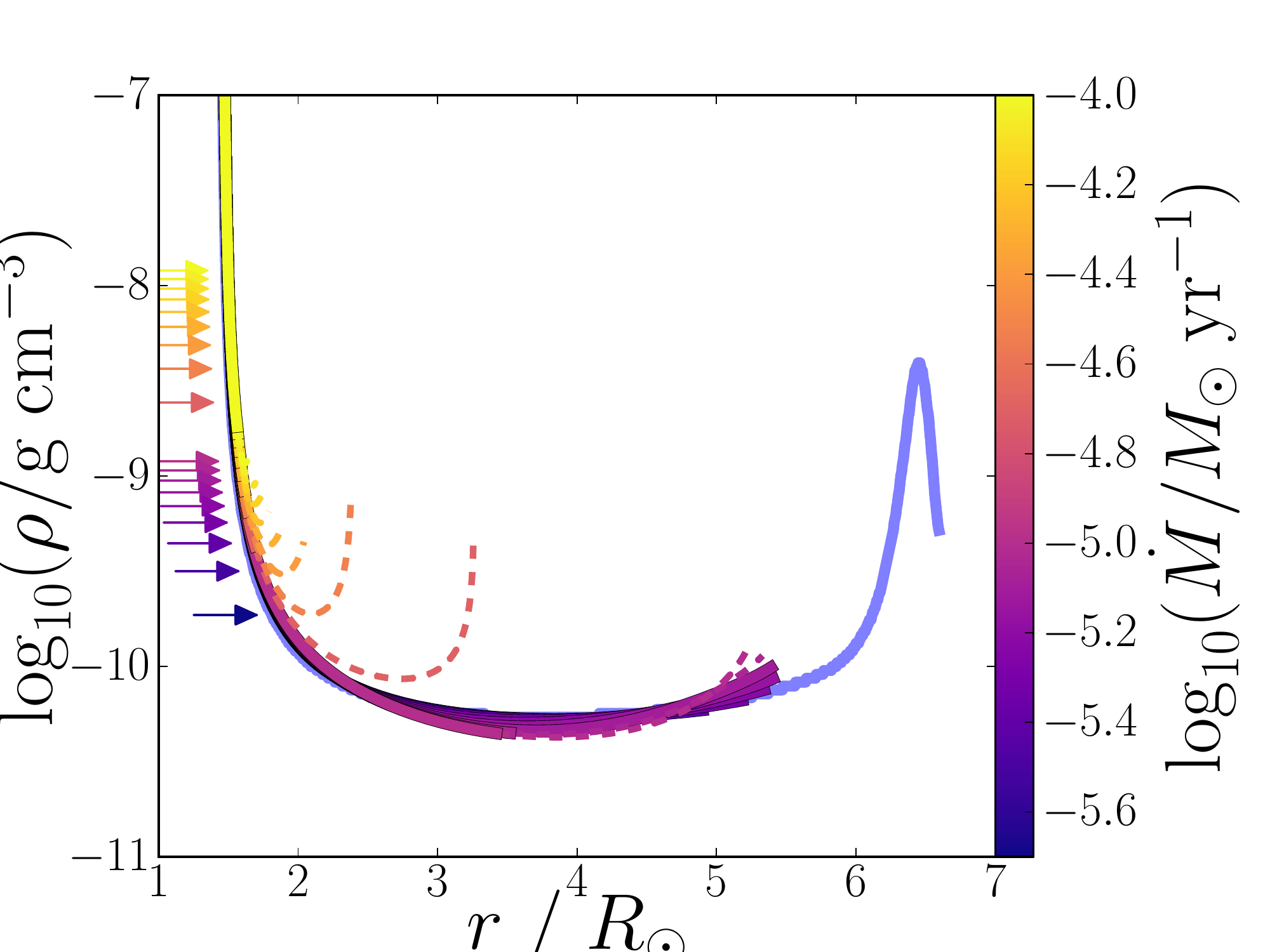}
    \caption{Stellar density profiles. Light blue is by \citet{grafener}, and the remaining are wind solutions across a range of mass loss rates. Arrows indicate the sonic point location for each wind model. Dashed regions indicate where the Rosseland approximation is no longer valid. }
    \label{fig:mesa_grafener_dens}
\end{figure}

\begin{figure}[h]
    \centering
    \includegraphics[width=\columnwidth]{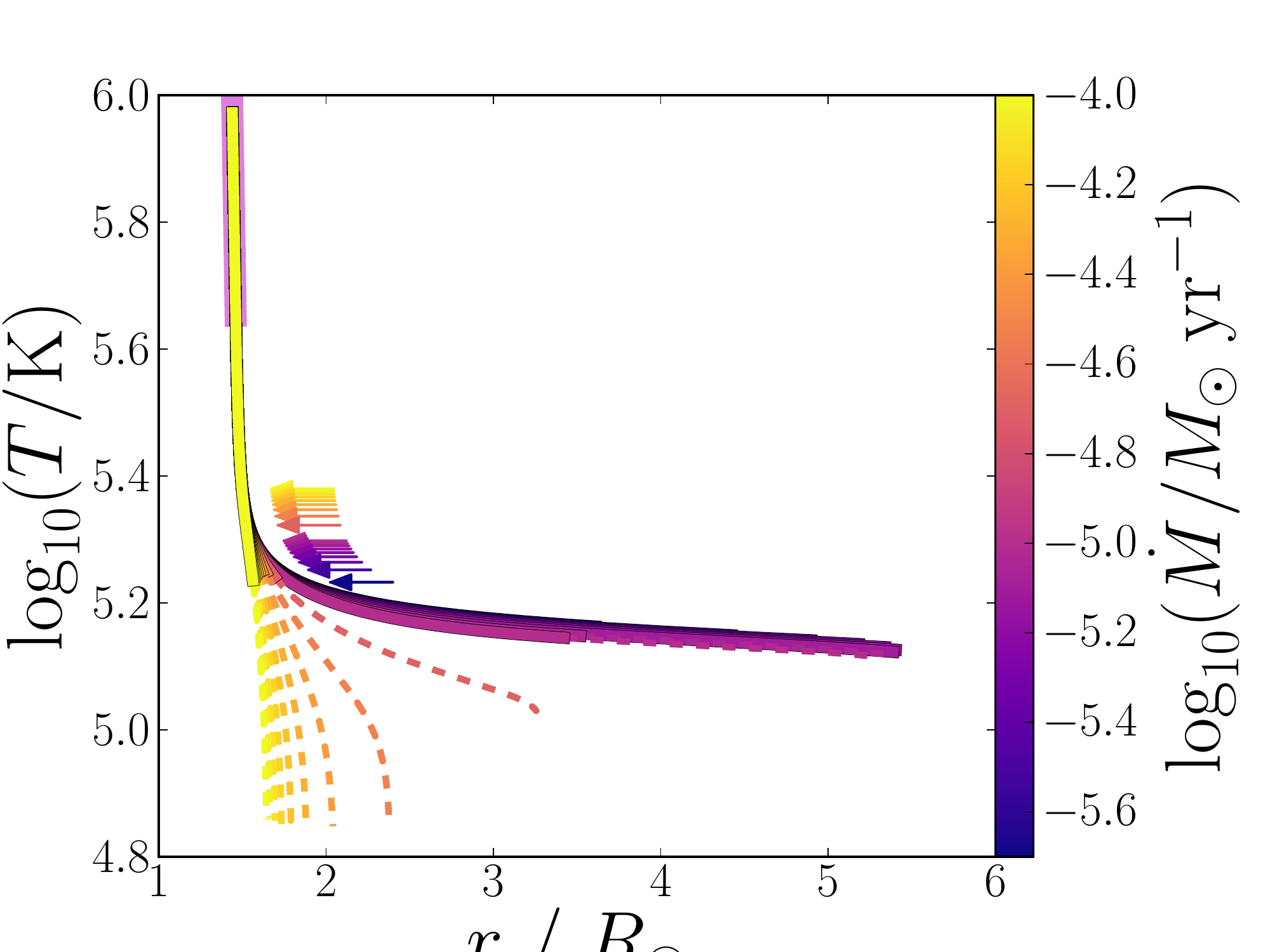}
    \caption{Temperature profile for the stellar wind solutions. Arrows indicate the sonic point location. Dashed regions indicate where the diffusion approximation is no longer valid. Note the contrast in temperature scale height between weak and strong winds. Compact wind models are truncated to temperatures above the partial ionization zone of helium ($10^{4.8}$\,K). }
    \label{fig:mesa_grafener_temp}
\end{figure}

Hydrostatic density profiles from MESA and \cite{grafener} are presented in Figure \ref{fig:mesa_grafener_dens}. The wind models converge towards the MESA solution in density and temperature (see Fig. \ref{fig:mesa_grafener_temp}) for $\rho>10^{-9}$\,g\,cm$^{-3}$ and $T>10^{5.2}$\,K. We note that wind solutions cannot converge exactly to the hydrostatic solution as the mass loss rate is not zero. However, the differences between all wind and hydrostatic solutions are proportional to the velocity, which %are found to rapidly diminish 
diminishes rapidly towards the interior. 

\begin{figure}[h]
    \centering
    \includegraphics[width=\columnwidth]{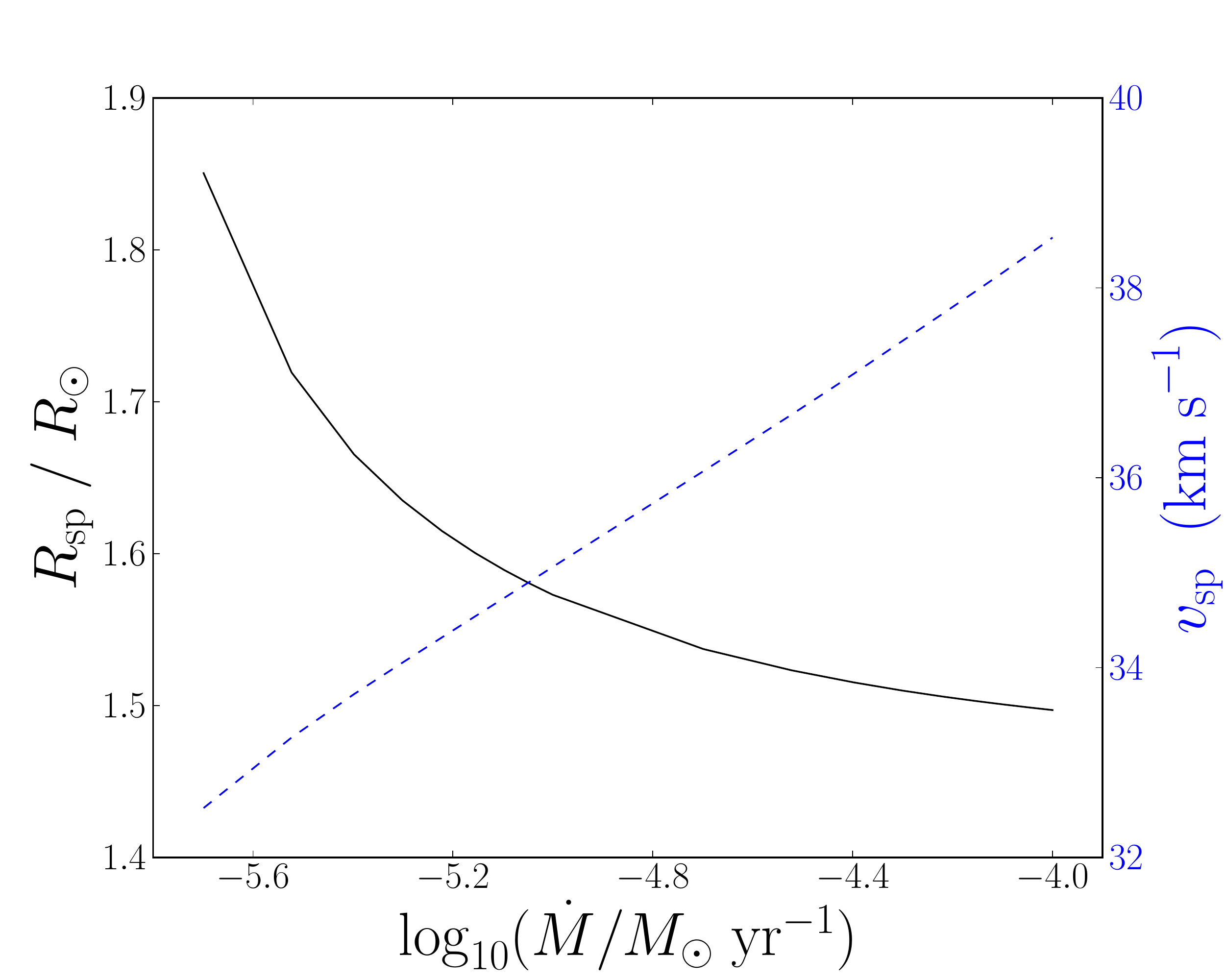}
    \caption{Radius and local sound speed at the sonic point location across a range of mass loss rates. The bifurcation in behaviour does not appear in the sonic point location or velocity.
    }
    \label{fig:mdot}
\end{figure}

A bifurcation of wind models exists across $\mdot_b = 2\times10^{-5} $M$_\odot$\,yr$^{-1}$. Winds with $\mdot>\mdot_b$ rapidly decline in temperature and do not extend far beyond one stellar radius before dropping to temperatures and optical depths outside our range of validity.  We refer to these as `compact winds'. In Figure \ref{fig:mdot}, we see that the sonic point occurs deeper within the star for increasing mass loss rates; however $R_\sonic$ varies smoothly with $\dot M$, so this is not the direct cause of the bifurcation.

\begin{figure}[h]
    \centering
    \includegraphics[width=\columnwidth]{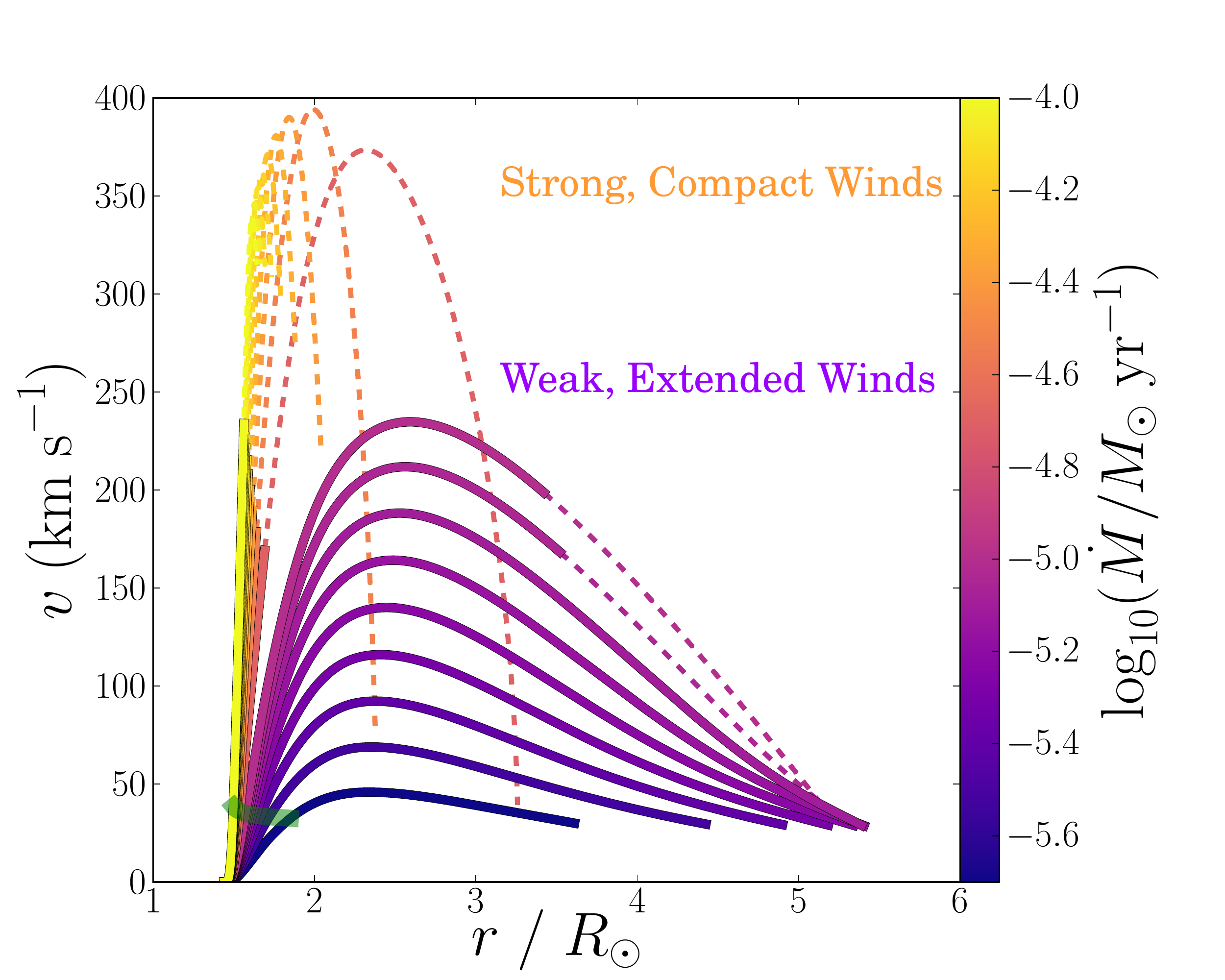}
    \caption{Wind velocity showing the bifurcation between strong, compact winds and weak, extended ones. The green line indicates the sonic point location and velocity, and dashed regions are where the Rosseland approximation becomes invalid. Note that the escape speed is $2100 (2 $R$_\odot/r)^{1/2}$km\,s$^{-1}$ for the 23 M$_{\odot}$.}
    \label{fig:mesa_grafener_vel}
\end{figure}

Weaker winds ($\mdot<\mdot_b$) are shallow in density and nearly isothermal throughout.  We refer to these as `extended winds'. Their structure strongly resembles the hydrostatic models with envelope inflation. While they are not hydrostatic envelopes, as this zone is outside the sonic point, it is plausible that they connect smoothly to the hydrostatic solution in the case $\mdot\rightarrow 0$. The peak speeds of these weak, extended winds are lower than those of the strong, compact winds (Figure \ref{fig:mesa_grafener_vel}) .

Figure \ref{fig:mesa_grafener_vel} presents the structure of $v$ and $c_i$.  All winds reach peak velocity with a maximum of $v\simeq400 \mrm{km \ s}^{-1}$, a factor of 
five slower than the local escape velocity.
Thus, no wind solutions are found to escape from radiation pressure alone for this star.

The Rosseland approximation is valid throughout our calculation of the structure of weak, extended winds with $\mdot<10^{-5.2}\,$M$_\odot$\,yr$^{-1}$.  These winds fail to 
reach escape velocity as they cross the Fe opacity bump, and rapidly decelerate at lower temperatures and larger radii.  We conclude that these cannot be the interior to a successful wind solution. 
Winds with higher mass-loss rates, especially the strong, compact branch of solutions, exit the regime of validity of our Rosseland approximation. Because Doppler enhancement of the line opacities becomes strong, these are candidates for successful wind solutions.

We postpone our explanation of the bifurcation in wind models until \S~\ref{sec:inflation}, where we shall analyze wind and envelope structures in the space of density and temperature.

\subsection{ Other helium stars} \label{SS:Various-masses}
To extend our modelling to helium stars of other masses, we rely on the
empirical relations of \cite{1992A&A...263..129S} to supply the inner boundary conditions. 

The stellar luminosity relation is
\beq
\mrm{log_{10}} \frac{L_*}{L_\odot} = 3.03 + 2.70 \left( \mrm{log}_{10} \frac{M_*}{\mrm{M}_{\odot}} \right) - 0.46\left( \mrm{log}_{10} \frac{M_*}{\mrm{M}_{\odot}}\right)^2,
\label{emasslum}
\eeq
which is accurate up to $\pm$0.1 dex for stellar masses between $3 \lesssim M_*/$M$_{\odot} \lesssim 65$. The luminosity for a 23 M$_\odot$ star from equation (\ref{emasslum}) is $\mrm{log}\left(L_*/L_\odot \right)=5.85$, and $\mrm{log}\left(L_*/L_\odot \right)=5.80$ from MESA. The luminosity at the base of the wind is also artificially adjusted such that the diffusive luminosity matches the luminosity from \cite{1992A&A...263..129S}. This correction is at most $10\%$ across all stellar masses and increases proportionally with mass loss rate.

It is important to ensure the temperature and radius are consistent at the inner boundary. The hydrostatic radius relation found for a WR star is
\beq
\mrm{log_{10}} \frac{R_*}{\mrm{R}_{\odot}} = -0.66 + 0.58\left( \mrm{log_{10}} \frac{M_*}{\mrm{M}_{\odot}} \right), 
\label{emassradius}
\eeq
which is accurate to $\pm$0.05 dex. However, only the corresponding `surface' temperature relation from the Stefan-Boltzmann law is available at this location. This temperature is not suitable as the surface is not freely emitting. 

In the MESA-generated 23 M$_{\odot}$ model, the temperature decreases by over an order of magnitude ($10^{5.4}$ K from $10^{6.8}$ K) across $10\%$ of the outer stellar envelope. We fix the temperature $\mrm{log}\left(T_0/\mrm{K}\right)= 5.8$ at the inner boundary and construct three sets of stellar wind models with varying base radii of $R_0=(1, 1.05, 1.10)\times R_*$. This serves to ensure the base of the wind is in close proximity to the true stellar conditions and determine how significantly the wind structure is affected by the depth of the potential well.  

%Present are 
We present density, temperature and velocity profiles (fig. \ref{fig:helium_dens}, \ref{fig:helium_temp}, and \ref{fig:helium_vel}) for $R_0=1.05 R_*$. We find the difference in base radii to only affect the critical mass loss rate for extended winds and the onset of line-force amplification. In general, the results are qualitatively similar to the 23 M$_\odot$ example. 

With increasing stellar mass, the extended wind models grow to larger radii and reach higher peak velocities. This structure is also more resilient to mass loss with increasing stellar mass. We find the critical mass loss rate for wind bifurcation scales approximately as $\mdot_b\propto M_*^2$. We found no extended wind solutions in stars less massive than 14 M$_\odot$. We discuss the physical origins of this limit further in Section \ref{sec:onset}.

Particular wind models for $M_*\ge30$ M$_\odot$ can approach the escape speed from radiation pressure, within the regime of validity of the Rosseland approximation and without line-force amplification. Line-force amplification, however, becomes important within a narrow range of wind velocities (150-200 km/s) for all stellar masses. 

\captionsetup[figure]{width=1\textwidth}
\begin{figure*}
  \centering
\begingroup
\captionsetup[subfigure]{width=1\textwidth}
 \subfloat[Density profiles of winds with strong, weak, and critical mass loss rates. The intermediate model shown bifurcates the weak, extended and strong, compact winds. All stars shown are capable of forming an extended wind. The radius extension grows with $L_*/M_*$, or stellar mass.]{\label{fig:helium_dens}\includegraphics[width=0.6\textwidth]{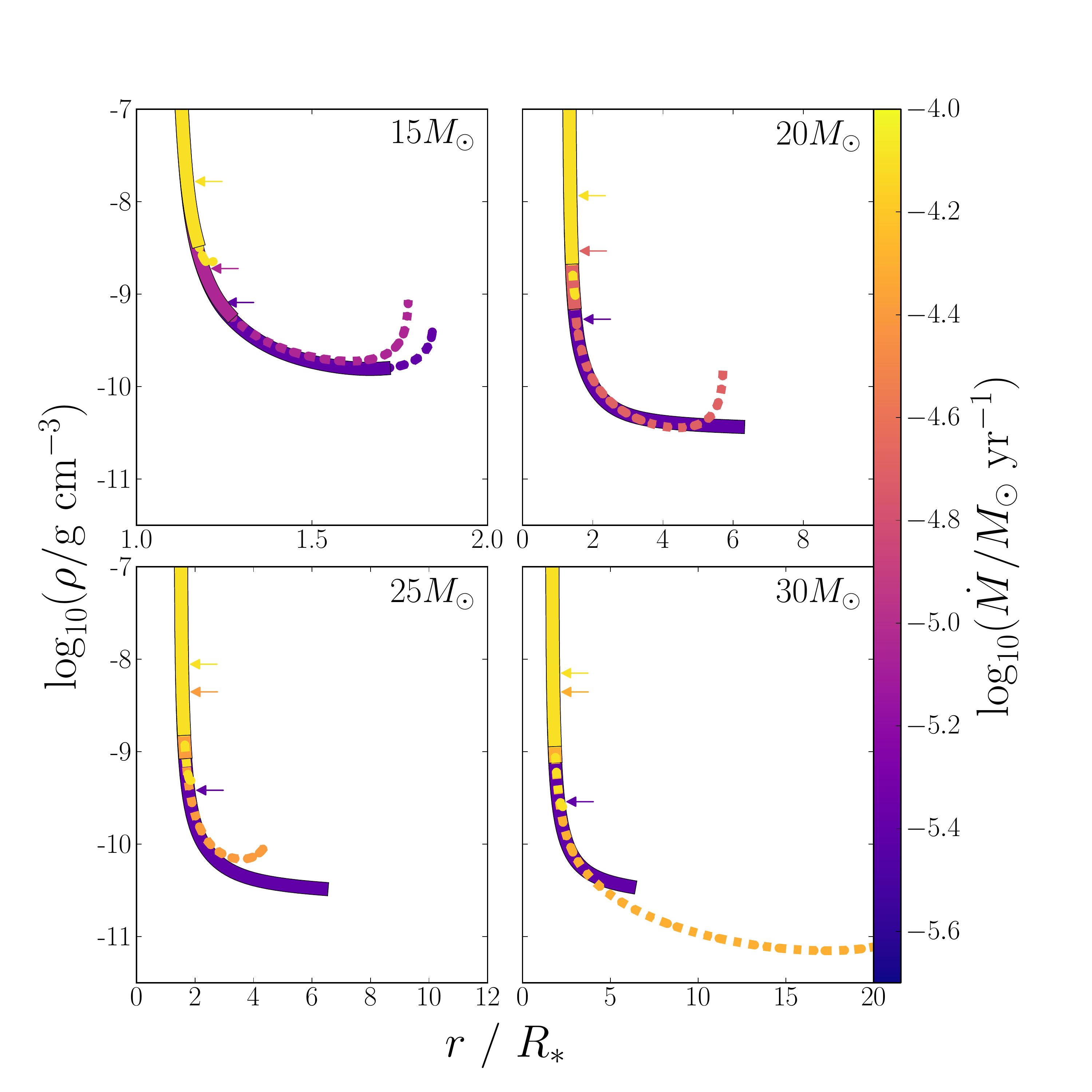}}\\
  \subfloat[Temperature profiles of winds with strong, weak, and critical mass loss rates. Strong winds are arbitrarily truncated at $T=10^{4.8}$\,K near the partial ionization zone of helium. Extended winds do not cool effectively and become practically isothermal. ]{\label{fig:helium_temp}\includegraphics[width=0.6\textwidth]{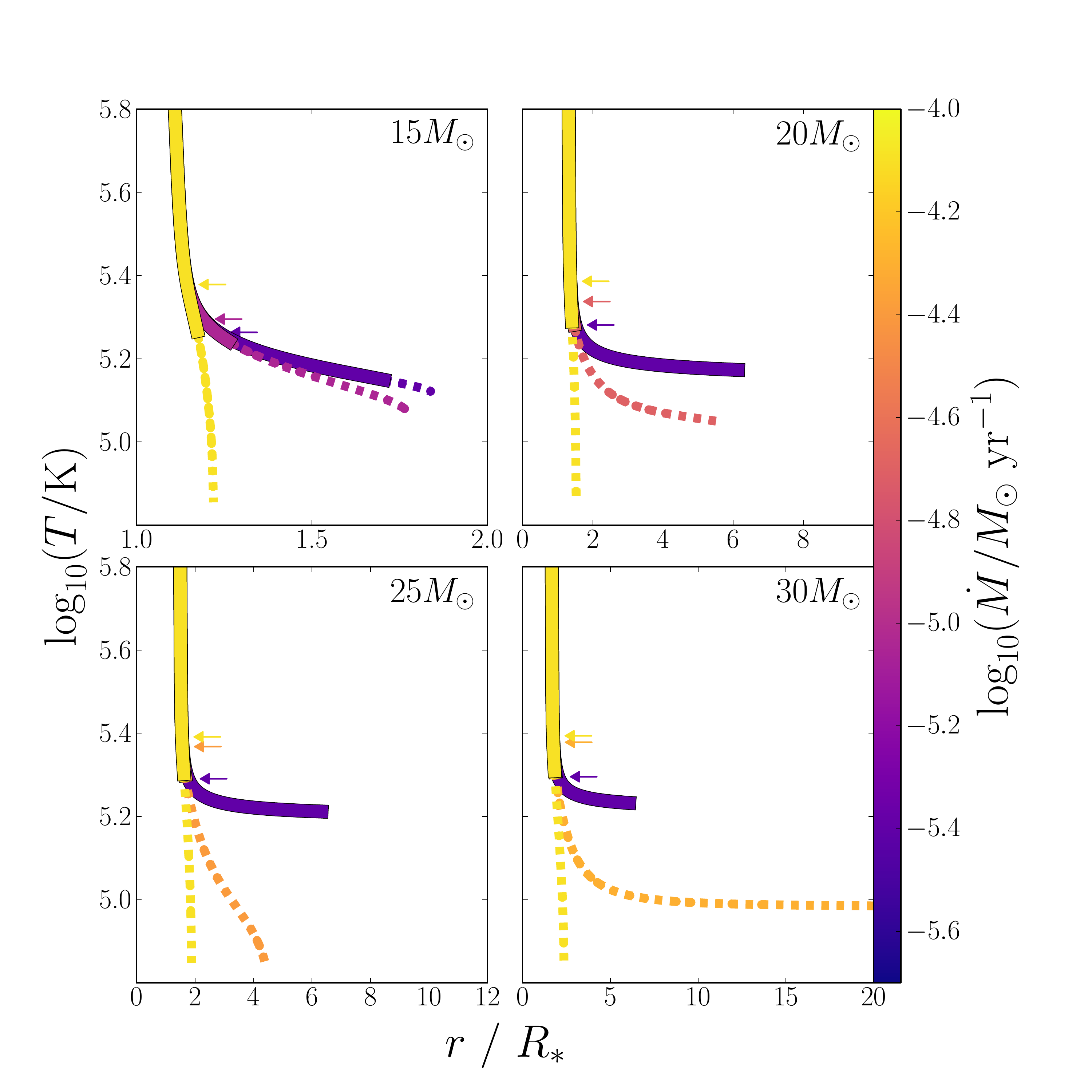}}
\endgroup
\caption[]{Profiles of stellar wind models for $M=15,20,25,30 $M$_{\odot}$ helium stars. For figures (a) and (b), arrows indicate the respective sonic point location. For all figures,  dashed regions indicate where the Rosseland approximation is invalid.}
\end{figure*}
\begin{figure*}
  \ContinuedFloat
  \centering

 \subfloat[Velocity profiles for all mass loss rates.  A green line indicates the sonic point location and velocity, and the black line is the local Kepler speed $v_k=\sqrt{GM_*/r}$. Extended winds are found to always become supersonic, if the driving is by radiation pressure alone. Note that for higher stellar masses, radiation pressure alone is capable of accelerating the wind to near escape speeds. Simulations of higher mass stars $M_*>30 $M$_{\odot}$ not presented here support this.] {\label{fig:helium_vel}\includegraphics[width=1\textwidth]{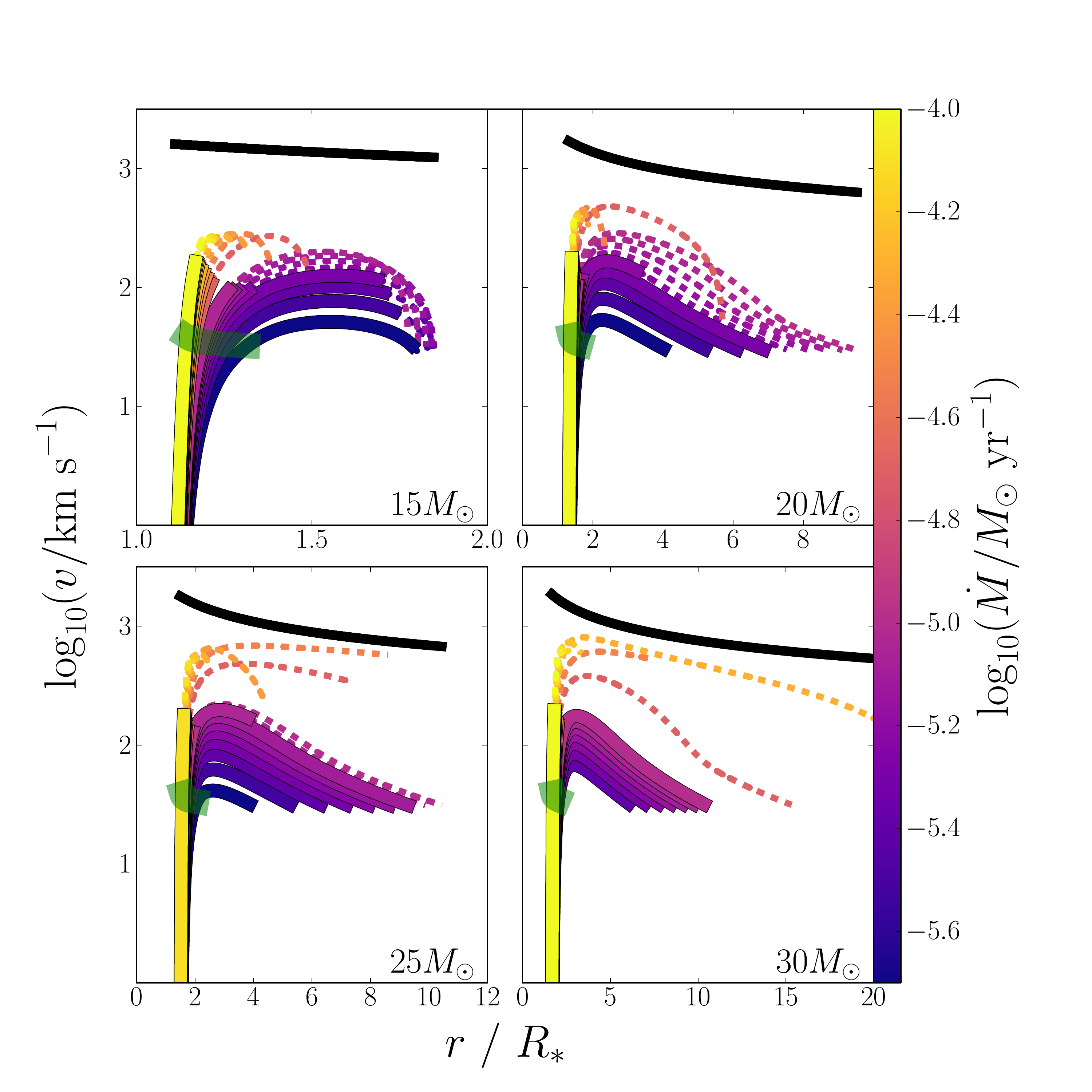}}
\caption[]{Profiles of stellar wind models for $M=15,20,25,30 $M$_{\odot}$ helium stars. For figures (a) and (b), arrows indicate the respective sonic point location. For all figures,  dashed-lines indicate where the Rosseland approximation is invalid.}
\end{figure*}

\begin{figure*}
    \includegraphics[width=1\textwidth]{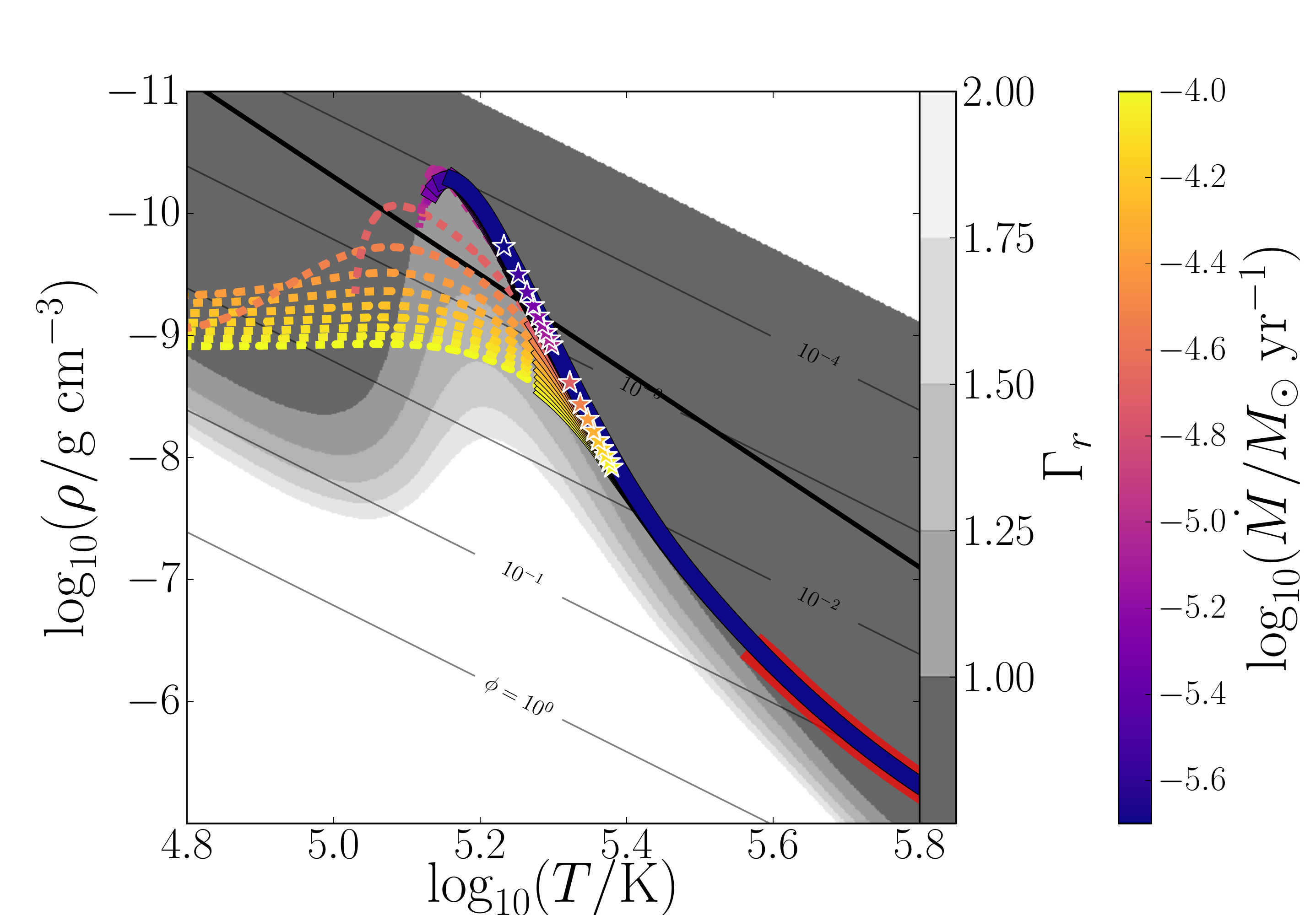}
    \caption{Shown is a diagram illustrating the onset of envelope inflation and extended winds for a 23 M$_{\odot}$ (or $L_*/M_*=2.7\times10^{4}(L_{\odot}/$M$_{\odot})$) star. The density and temperature structure of the MESA-generated model (thick red line) and stellar wind (colours) extends from the stellar interior (bottom-right) to the surface (top-left). 
The stellar wind models trace the $L_*\kappa(\rho,T)/(4\pi c G M_*)=1$ (greyscale) contour and become strongly radiation-dominated (decreasing $\phi=P_g/(4P_r)$) until reaching the sonic point (stars). Wind models that cross the line of inflation (thick-black) are extended in radius, as the temperature scale height and stellar radius become comparable (see eq.\ \ref{e_H_T}). Dashed regions of the wind models indicate where the Rosseland approximation is no longer valid.  A 14.25 M$_{\odot}$ star with marginally extended winds is shown in Fig. \ref{fig:m14}. }
    \label{fig:m23}
\end{figure*}

\section{Discussion: Inflation, Inversion, and Stability}
\label{sec:inflation}

We begin by dividing $dP/dr$ (obtained from the momentum equation, eq.~\ref{emom}) by $dP_r/dr$ (from the diffusion equation, eq.~\ref{ediff}) to obtain 
 
\beq
\frac{\dd P}{\dd P_r} = \left(1 + \frac{v^2v'}{v_k^2}\right)\Gamma_r^{-1}.
 \label{estructure}
\eeq

Equation (\ref{estructure}) provides important information about structure and stability, especially in the context of a radiation-dominated outer WR envelope and wind, where $M(r)/L(r) \simeq M_*/L_*$ so that the Eddington factor $\Gamma_r\simeq \kappa L_*/(4\pi GM_* c) $ is almost exactly proportional to $\kappa$.    (Indeed, $M(r)$ is almost constant in our solutions and $L(r)$ varies by at most 10\%.)   Furthermore $\kappa$ is a function of density and temperature, at least where the Rosseland approximation is valid.   Finally, the inertial term $v^2 v'/v_k^2$ is negligible in subsonic regions, where the flow is nearly hydrostatic, but becomes important in supersonic regions.  However,  we saw in Equation (\ref{e_Gamma_sp}) that $\Gamma_r$ takes a specific value, very close to unity, at the sonic point.   Therefore, the structure of the outer stellar envelope, and the transition to a wind, can be related directly to the opacity law in the plane of density and temperature. 

\subsection{Convective instability} \label{SS:convection} 

The criterion for convective instability, equation (\ref{e_ConvInstab}), can also be assessed within this plane.  From equation (\ref{estructure}) and the relation $\nabla^{-1} = 4(1-\beta)\dd P/\dd P_r $, along with the expression for $\nabla_{\rm ad}(\beta)$ in a monatomic gas (\cite{1990sse..book.....K}; Eq. (13.21)), we find that the flow is unstable ($\nabla_{\mrm{rad}} \ge \nabla_{\rm ad}$) where 

\beq \label{e_ConvInstabEvaluated} 
\Gamma_r \ge \left(1 + \frac{v^2v'}{v_k^2}\right)\Gamma_c,
\eeq 
with 
\beq \label{egamma_convection}
\Gamma_c \equiv{8(4-3\beta)(1-\beta) \over 8(4-3\beta) - 3 \beta^2} < 1.  
\eeq

\subsubsection{Outflows inhibit convection} \label{SSS:OutflowsInhibitConv}

In the absence of any wind ($v=0$), this criterion sets a very specific value of  $\Gamma_r=\Gamma_c(\beta)$ above which an envelope is unstable -- effectively dividing the phase space into stable and unstable regions of $\rho$ and $T$ (for hydrostatic models). This instability condition is necessary but not always sufficient: an accelerating wind, with $v\neq0$ and $v'>0$, is more stable on account of the inertial term $v^2 v'/v_k^2$ in equation (\ref{e_ConvInstabEvaluated}).  It is therefore necessary to examine in some detail the convective instability at the sonic point. 

Evaluating equation (\ref{e_ConvInstabEvaluated}) provides a condition for convective stability at the sonic point, valid where $\phi,q_i\ll1$:
\beq \label{e_convection_sp}
7\phi + k_{\rho} + k_T/3 < 0.
\eeq
Here $\phi$ and $k_{\rho}$ are positive, but the sonic point forms at temperatures somewhat above $T\simeq10^{5.2}$\,K where $k_T$ is sufficiently negative. As a result, we find that all the stellar wind models in this paper are stable against convection throughout their subsonic regions.

We note that \cite{2009A&A...499..279C} attribute WR star variability to convection driven by the iron opacity peak, on the basis that convection should set in near the wind sonic point.   However our finding that the subsonic region and sonic point are stably stratified indicates that radiation-driven acoustic instabilities \citep{2003ApJ...596..509B} are a more likely cause.

This does not imply, of course, that hydrostatic envelopes, or envelopes with very low mass-loss rates, cannot contain convective regions.  These envelopes pass through $\Gamma_{r,\sonic}$ at subsonic speeds, rather than crossing a sonic point.   

\subsubsection{Hydrostatic models convect} 

Indeed, convection appears to be inevitable for radiation-dominated hydrostatic envelopes interacting with the Fe or He peak, as a consequence of equation (\ref{estructure}) with $v=0$, rewritten  as $dP_g/dP = 1-\Gamma_r$;
in words, gas pressure declines outward when $\Gamma_r<1$.

It is impossible for a hydrostatic, low-$\beta$ envelope to exist without convection in the presence of an opacity peak.  Consider Figure \ref{fig:m23} or \citet{grafener}'s Figure 5, which plot $\Gamma_r$ in the space of $\rho$ and $T$ or $P$ and $P_r$.  Following a solution outward to decreasing $P$ and $T$, the density drops dramatically to skirt the hot side of the $\Gamma_r>1$ zone.   In the process, $\Gamma_r$ self-consistently becomes very close to unity, because $dP_g/dP_r$ is very small along the $\Gamma_r=1$ contour.  On the cold side of the bump, however, $\rho$ and $P_g$ increase rapidly along this contour (the density inversion).  To remain close to this contour requires $P_g$ to rise, which requires $\Gamma_r>1$; but this implies that $\Gamma_r$ passed through the convective threshold ($\Gamma_r=\Gamma_c<1$) along the way.

\subsection{Onset of Envelope Inflation and Extended Winds}
\label{sec:onset}

The weak, `extended' winds closely follow the $\Gamma_r=1$ contour, but do so by becoming supersonic as they cross the opacity peak, and subsonic once again as they exit it. Strong, `compact' winds, on the other hand, traverse the opacity peak more directly, plunging deep within the super-Eddington region. It is the inertial term which allows a wind to enter the opacity peak without developing a gas pressure or density inversion $\Gamma_r=(1+v^2v'/v_k^2)(1-dP_g/dP)$.

What is the underlying cause of the bifurcation in wind behaviour?  A major clue is that the bifurcation coincides with the dashed line on Figure \ref{fig:m23}, which denotes the condition $q_r=1/2$, or $2 c_r = v_k$.  (In the plot, $v_k$ is evaluated at the base of the wind $R_*\sim 1.44 $R$_{\odot}$.)  The importance of this condition arises from fact that the temperature scale height $H_T = |\dd r/\dd \ln T|$ can be evaluated, in any region governed by the radiation diffusion equation (eq.~\ref{erad}), as 
\beq \label{e_H_T} 
H_T =\left(2 c_r \over v_k\right)^2  \Gamma_r^{-1} r. 
\eeq 
Envelopes and winds that follow the $\Gamma_r=1$ contour will contain an extended temperature plateau in which $H_T>r$. The result is an inflated envelope or a specimen of our weak, `extended' wind class. On the other hand, if the 
density is sufficiently high 
the line of inflation is avoided. This causes the temperature to plummet through the opacity peak, with $H_T/r$ decreasing further as $\Gamma_r$ exceeds unity; the result is a strong, `compact' wind that accelerates rapidly.  

Importantly, this criterion depends only on the validity of the radiation diffusion equation, so it is equally valid within hydrostatic models as in winds; this explains why the weak, extended winds track the profile of the hydrostatic model.

The wind model that traces the line of inflation through the opacity peak separates the weak `extended' and strong `compact' winds. We select the conditions where the line of inflation ($2 c_r = v_k$) and opacity peak ($T_p=10^{5.2}$\,K) intersect to estimate the bifurcating mass loss rate, 
\beq \label{emdotb}
\mdot_b = 4\pi r_p^2 \rho_p v_p =4 \pi R_*^2 \left( \frac{3GM_*}{4aR_*T_p^4}\right) v_p.
\eeq
The location of the opacity peak $r_p$ is approximately at the base of the wind $R_*$. Since the pressure gradient follows the line of inflation ($dP/dP_r=1+dP_g/dP_r \simeq 1 + 5\phi$), equation (\ref{evprime}) and (\ref{estructure}) supply a local estimate for the velocity 
\beq \label{evc}
\frac{v_p^2}{v_p^2-c_i^2} = \frac{(1+5\phi)\Gamma_r - 1}{q_i + \Gamma_r \left(1+\phi\right) - 1}. 
\eeq
Finally, the Eddington factor $\Gamma_r=\kappa(\rho_p,T_p)L_*/(4\pi c GM_*)$ is calculated with the OPAL opacity table and stellar luminosity relation equation (\ref{emasslum}). 

\captionsetup[figure]{width=1\columnwidth }
\begin{figure}
    \centering
    \includegraphics[width=\columnwidth]{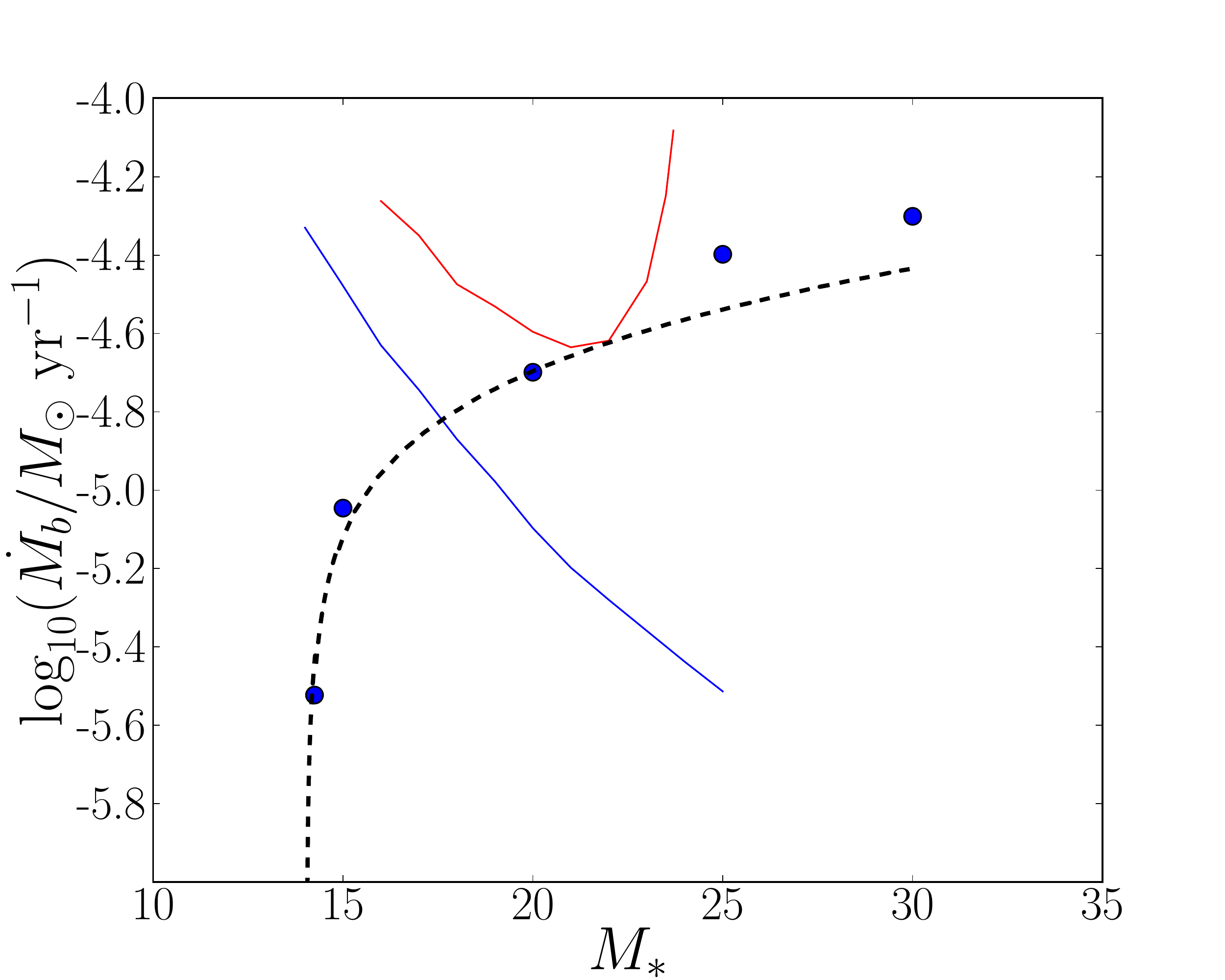}
    \caption{The numerical (points) and analytic (dashed line) estimates for the critical mass loss rate for bifurcation across stellar mass. For mass loss rates below this boundary, we find slow, extended solutions that are incompatible with successful WR winds. With our assumptions, the analytic estimate for $\mdot_b$ converges to zero at $M_*=14.04\ $M$_{\odot}$. The analytic model underestimates the critical mass loss rate by a small value towards more massive stars as the location of the iron opacity peak is further than the base of the wind. See eq.\ (\ref{emdotb}). Blue and red lines are estimates from \cite{petrovic} and \cite{grafener}, respectively.}
    \label{fig:mdotb}
\end{figure}
\captionsetup[figure]{width=1\textwidth }

Although our estimate of $\mdot_b$ is approximate, we find excellent agreement with the wind models generated from the sequence of helium stars (see Figure \ref{fig:mdotb}). For the 23 M$_{\odot}$ case generated with MESA, we predict $\mdot_b = 1.6\times10^{-5}$ M$_\odot$\,yr$^{-1}$, which is in excellent agreement with our numerical results ($2\times10^{-5} $M$_\odot$\,yr$^{-1}$) as well. $\mdot_b$ is marginally underestimated towards more massive stars since the location of the opacity peak is further from the base of the wind (ie. $r_p>R_*$).

\begin{figure*}
	\centering
    \includegraphics[width=\textwidth]{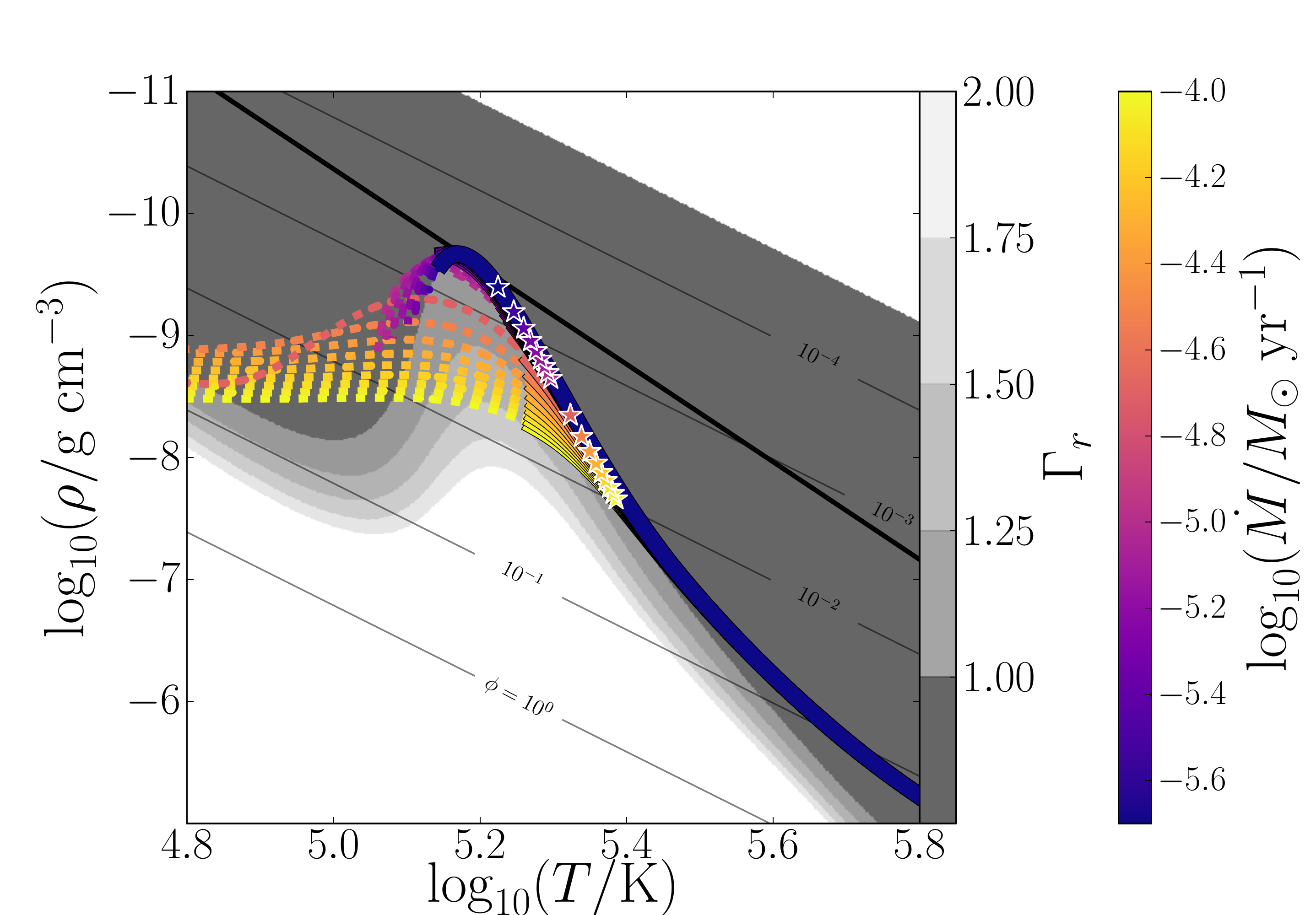}
    \caption{ A structural diagram of the stellar winds for a $14.25 $M$_{\odot}$ (or $L_*/M_*=2.4\times10^4L_{\odot}/$M$_{\odot}$) star. The $\Gamma_r=1$ contour recedes to higher densities and away from the line of inflation for stars with decreasing $L_*/M_*$ or stellar mass (see eq.\ (\ref{emasslum})). At $14.04 $M$_{\odot}$, the $\Gamma_r=1$ contour and line of inflation intersect at one point, and any non-zero mass loss rate is sufficient to prevent the formation of an extended wind (see eq.\ (\ref{emdotb}) and (\ref{evc}) for the exact criterion). Therefore, the critical mass loss rate for bifurcation increases with stellar mass. }
    \label{fig:m14}
\end{figure*}

$\mdot_b$ is found to rapidly decline towards lower stellar mass. This is confirmed with an additional set of stellar wind models constructed for a $M_*=14.25 $M$_{\odot}$ helium star. The reason is seen from Figure \ref{fig:m14} which displays the star on a $\rho$ and $T$ plane. In comparison to Figure \ref{fig:m23}, the $\Gamma_r=1$ contour recedes to higher densities for stars with lower mass or $L_*/M_*$. This reduces the $\mdot_b$ necessary to avoid crossing the line of inflation. Higher mass stars can generate winds that extend tens of stellar radii and, likewise, increases the $\mdot_b$ necessory to erase the structure.

At exactly $M_*=14.04 $M$_{\odot}$, the line of inflation and $\Gamma_r=(1+5\phi)^{-1}\simeq1$) contour intersect at one point, and any non-zero mass loss rate will form a compact wind. Therefore, we find a minimum stellar mass or $L_*/M_*$ for envelope inflation from the iron opacity bump. 

\cite{petrovic} present a different argument for approximating $\mdot_b$. They state the inflated envelope is preserved, if the inertial term is smaller than gravitational acceleration (ie. $v<\sqrt{GM_*/R_*}$). Evaluating $\mdot_b$ at the hydrostatic radius (Eq. \ref{emassradius}) and the minimum envelope density $\rho=\rho_{\mrm{min}}$, as prescribed by \cite{petrovic}, generates the blue line in Fig. \ref{fig:mdotb}. Since inflated envelopes trace the $\Gamma_r=1$ contour, we estimate $\rho_{\mrm{min}}$ at the opacity peak (see Fig. \ref{fig:m23}). We use Table 1 from \cite{grafener} to generate the red line in Fig. \ref{fig:mdotb}. This is a similar approximation except $\mdot_b$ is reduced by 0.4 dex and evaluated at the envelope density minima location (see Eq. (33) from \cite{grafener}). These approximations suggest an inflated envelope becomes more robust for lower stellar mass, which is not in agreement with our models.

\subsection{The nature of weak WR winds}
\label{sec:wind_fail}

We have found that stars with mass loss rates below $\mdot_b$ do not, within steady, spherically symmetric models, maintain the strong compact winds that we have identified as good candidates for WR winds. For mass loss rates below this limit, we find weak, extended wind solutions ​that fail to launch winds by the iron opacity bump, because the Rosseland approximation remains valid as they decelerate. We infer these `winds' fall back onto themselves, unless they are able to reach the helium opacity bump.

We note that extended envelopes in which $2 c_r>v_k$,  $\Gamma_r\simeq 1$, and $\beta\ll1$ are formally unbound, in the sense that they have a positive Bernoulli parameter (${\cal B} > 0$ in equation (\ref{enrg})). However this is not relevant to the bifurcation in wind models, because diffusion is rapid enough that radiation is not trapped in WR winds.

\subsection{Radiation-driven acoustic instabilities}\label{SS:acoustic-instabilities} 
The acoustic instability identified by \cite{2003ApJ...596..509B}
is a source of effects not accommodated within our models. 
For a pure helium WR star with solar metallicity, the instability occurs for log $T/$K $\lesssim 5.7$, which is achieved at radii beneath the sonic point. At log $T/$K $=5.7$, \citeauthor{2003ApJ...596..509B} identify a wavelength of fastest asymptotic growth approximately $\lambda_{\mrm{max}}/r=2\pi q_i/(\Gamma_r k_{\rho})\sim10^{-1}$, which exceeds the local pressure scale height $H_p/r\sim10^{-2.3}$. We hypothesize that growth is suppressed for wavelengths larger than the pressure scale height, and evaluate the growth rate for $\lambda = H_P$. The amplitude of this mode increases by 10 e-foldings from the point of instability to the sonic point. Given initial perturbations greater than $10^{-4}$, the instability will become nonlinear in the subsonic domain and alter the conditions for wind launching.

\cite{2015arXiv150905417J} perform three dimensional local radiation hydrodynamic simulations of an envelope patch at the iron opacity peak. They find the density inversions found in one dimensional simulations correspond to large-scale density fluctuations and supersonic turbulent velocity fields in three dimensions. Although the local simulations cannot determine whether a large-scale wind is initiated, the structural characteristics may be realized in Wolf-Rayet stars with weak extended winds. ​Density fluctuations may give rise to a clumped (or porous) atmosphere in which the effective opacity, and Eddington ratio, is modified and enhanced (or reduced). We anticipate that our analysis of outer WR envelopes and inner WR winds applies just as well to the modified opacity law as to the unmodified one. We direct the reader to \cite{grafener} and \cite{2013A&A...560A...6G} for the effects of clumping on the structure of an inflated envelope and opacity enhancement.

\section{Conclusions} \label{S:conclusions}
We draw several conclusions from our investigation of the transition from envelope to wind within WR stars.   

First, we find that the inflation of stellar envelopes, caused by the iron opacity peak and observed within hydrostatic models of WR winds, extends into a class of weak, `extended' winds. However, above the critical mass loss rate $\mdot_b$, these are replaced by a strong, `compact' class of solutions. Physically, this change in behavior arises from a change in the ratio of the temperature scale height $H_T$ to the local radius.    However our weak, extended winds fail to accelerate within the regime of validity of our Rosseland approximation.   In contrast the strong, compact branch is compatible with acceleration to escape speeds (outside the regime of the Rosseland approximation).  It is also compatible with the observed mass loss rates of WR stars. 

Second, we find that continuum-driven WR winds are always convectively stable at the sonic point.  Within a hydrostatic envelope, convection sets in at a critical Eddington factor $\Gamma_r$ that is slightly higher than the value of $\Gamma_r$ at the wind sonic point; in a moving envelope, an inertial term raises this critical value further.   Since the Eddington factor is increasing through the sonic point, the sonic point is always reached prior to the onset of convection (if it is reached at all). 

Third, our adoption of the Rosseland approximation limits the applicability of our results in two ways.   At large radii (usually outside the sonic point), our approach becomes invalid; the effective opacity is higher than the Rosseland mean, due to Doppler effects.   We nevertheless probe the envelope-wind transition for a variety of mass loss rates in order to identify solutions compatible with the formation of a wind in the Doppler-enhanced regions.   However, we also neglect acoustic instabilities that set in below the sonic point and may grow sufficiently to suppress the effective opacity relative to the Rosseland mean.   While we do not predict the magnitude of this effect, we hypothesize that our analysis remains valid so long as the Rosseland opacity is replaced with the effective opacity. 

Finally, we note that our results are not restricted to WR star winds, but apply to any object with a sufficiently optically-thick, continuum-driven wind stimulated by an increase in the opacity.

\acknowledgments This work was supported by a Discovery Grant from NSERC, the Canadian National Sciences and Engineering Research Council.  SR is supported by a Alexander Graham Bell Canada Graduate Scholarship. We thank John Hillier, Norman Murray, Chris Thompson, Marten van Kerkwijk, and the anonymous referee for thoughtful comments and conversations.

\end{document}